\documentclass[12pt,reqno,a4paper]{article}
\usepackage[OT1]{fontenc}
\usepackage{authblk,amsthm,amsmath, amsfonts, caption, multirow, float, graphicx, tikz, pdfpages, setspace, enumitem, fancyhdr, longtable, accents} 
\usepackage[symbol]{footmisc}
\usepackage[misc]{ifsym}
\usepackage[normalem]{ulem}
\usepackage[round]{natbib}
\usetikzlibrary{arrows,shapes,shapes.geometric}
\usepackage{pdflscape} 
\usepackage[colorlinks,citecolor=blue,urlcolor=blue]{hyperref}
\usepackage[margin=2.5cm,headsep=10mm,footskip=15mm]{geometry}
\usepackage{xcolor}
\definecolor{green1}{rgb}{0.0, 0.5, 0.0}
\definecolor{chocolate}{rgb}{0.48, 0.25, 0.0}
\definecolor{dukeblue}{rgb}{0.0, 0.0, 0.61}
\colorlet{brown1}{brown!70!black}
\colorlet{blue1}{blue!70!black}
\colorlet{notgreen}{blue!50!yellow}
\usepackage{array}
\newcolumntype{L}[1]{>{\raggedright\let\newline\\\arraybackslash\hspace{0pt}}p{#1}}
\newcolumntype{X}[1]{>{\raggedright\let\newline\\\arraybackslash\hspace{0pt}}m{#1}}
\newcolumntype{C}[1]{>{\centering\let\newline\\\arraybackslash\hspace{0pt}}m{#1}}
\newcolumntype{R}[1]{>{\raggedleft\let\newline\\\arraybackslash\hspace{0pt}}m{#1}}

\usepackage{soul}
\usepackage[framemethod=tikz]{mdframed}

\mdfdefinestyle{myFigureBoxStyle}{backgroundcolor=black!5,, roundcorner=3pt,tikzsetting={draw=black, line width=1pt}}%

\makeatletter
\newcommand\fs@myRoundBox{\def\@fs@cfont{\bfseries}\let\@fs@capt\floatc@plain
	\def\@fs@pre{\begin{mdframed}[style=myFigureBoxStyle]}%
		\def\@fs@mid{\vspace{\abovecaptionskip}}%
		\def\@fs@post{\end{mdframed}}\let\@fs@iftopcapt\iffalse}
\makeatother

\theoremstyle{plain}

\newtheorem*{result*}{Result}

\theoremstyle{definition}

\newtheorem*{example*}{Example}

\newcommand{\N}{\ensuremath{\mathcal{N}}}

\newcommand{\V}{\ensuremath{\mathcal{V}}}
\newcommand{\s}{\ensuremath{\textbf{\emph{s}}}}

\newcommand{\z}{\ensuremath{\textbf{\emph{z}}}}

\newcommand{\Y}{\ensuremath{\textbf{\emph{Y}}}}

\newcommand{\W}{\ensuremath{\textbf{\emph{W}}}}

\newcommand{\x}{\ensuremath{\textbf{\emph{x}}}}

\newcommand{\logit}{\mbox{logit}}

\newcommand{\ttheta}{\ensuremath{\boldsymbol{\theta}}}

\newcommand{\pr}{\ensuremath{\mbox{Pr}}}

\newcommand{\tbf}[1]{\noindent\textbf{#1}}

\title{Modelling spatially autocorrelated detection probabilities in spatial capture-recapture using random effects} 

\author[1]{Soumen Dey\footnote{\tbf{E-mail}: \href{mailto:soumenstat89@gmail.com}{\textit{soumenstat89@gmail.com}} }}
\author[1]{Ehsan M. Moqanaki}
\author[1]{Cyril Milleret}
\author[1]{Pierre Dupont}
\author[1,2]{Mahdieh Tourani}
\author[1]{Richard Bischof}

\affil[1]{Faculty of Environmental Sciences and Natural Resource Management, Norwegian University of Life Sciences, 1432 $\mathring{\text{A}}$s, Norway}
\affil[2]{Department of Wildlife, Fish, and Conservation Biology, University of California, Davis, CA, USA}

\begin{document}
\tikzstyle{every picture}+=[remember picture] 
		
\everymath{\displaystyle}

\setlength{\abovedisplayskip}{5pt}
\setlength{\belowdisplayskip}{5pt}
\date{}
\maketitle

\tbf{Running headline}: Spatially autocorrelated  detectability in SCR


\vspace{10cm}
\pagebreak



	


{\footnotesize
\begin{center}	{\bf Abstract } 
\end{center}

Spatial capture-recapture (SCR) models are now widely used for estimating density from repeated individual spatial encounters. SCR accounts for the inherent spatial autocorrelation in individual detections by modelling detection probabilities as a function of distance between the detectors and individual activity centres. However, additional spatial heterogeneity in detection probability may still creep in due to environmental or sampling characteristics. if unaccounted for, such variation can lead to pronounced bias in population size estimates.

Using simulations, we describe and test three Bayesian SCR models that use generalized linear mixed models (GLMM) to account for latent heterogeneity in baseline detection probability across detectors using: independent random effects (RE), spatially autocorrelated random effects (SARE), and a two-group finite mixture model (FM).

Overall, SARE provided the least biased population size estimates (median RB: -9 -- 6\%). When spatial autocorrelation was high, SARE also performed best at predicting the spatial pattern of heterogeneity in detection probability. At intermediate levels of autocorrelation, spatially-explicit estimates of detection probability obtained with FM where more accurate than those generated by SARE and RE. In cases where the number of detections per detector is realistically low (at most 1), all GLMMs considered here may require dimension reduction of the random effects by pooling baseline detection probability parameters across neighboring detectors (``aggregation") to avoid over-parameterization.

The added complexity and computational overhead associated with SCR-GLMMs may only be justified in extreme cases of spatial heterogeneity. However, even in less extreme cases, detecting and estimating spatially heterogeneous detection probability may assist in planning or adjusting monitoring schemes.

\vspace{1cm}

\tbf{Keywords}: Spatial capture-recapture, Detection probability, Spatial autocorrelation, Generalised linear mixed model, Random effects, Finite mixture model, Population size estimation
} 



\pagebreak

\section{Introduction}\label{sec:intro}
Spatial capture-recapture (SCR) models are now widely used to estimate demographic parameters, particularly density. SCR data inherently varies across space because animal movements are not completely random and an individual is more likely to be detected close to its centre of activity (`activity centre', AC). SCR models account for and, in fact, exploit such spatial heterogeneity in detection by modelling the detection probability as a decreasing function of distance between a detector - e.g., an observer, a trap, or a search location - and a latent AC \citep{efford2004density, borchers2008spatially}.  
However, the relative distance between a detector and an AC may not be the only cause of variation in detection probability. Spatially variable and autocorrelated detection probability can occur due to various other factors, such as local differences in how animals use space and how sampling is performed \citep{moqanaki2021consequences,stevenson2021spatial}. 


Known sources of variation in detection probability are readily modelled in SCR using covariates, for example, through proxies or direct measures of sampling effort \citep{efford2013varying}, resource selection data obtained from telemetry studies \citep{royle2013integrating}, or information about landscape connectivity \citep{sutherland2015modelling}. However, not all sources of variation are known and fully observed. For example, local site-specific characteristics affecting detector exposure, or effect of local atmospheric conditions on the genotyping success rate of non-invasively collected DNA samples may remain unaccounted for during SCR analyses \citep{moqanaki2021consequences, kendall2019bear, efford2013varying}. Furthermore, large-scale wildlife monitoring programs sometimes include both structured and unstructured sampling data. The latter may be data  collected by the general public to increase the extent and/or intensity of sampling \citep{thompson2012framework, bischof2020estimating}. Unstructured and opportunistic sampling data is likely to be associated with unknown spatial variation in detection probability. Unmodelled spatial variation in detection probability, particularly in the presence of high spatial autocorrelation, can lead to biased and overdispersed population size estimates in SCR analyses \citep{moqanaki2021consequences}. A worst-case scenario are pockets or clusters of detectors where, unbeknownst to the investigator, detection probability is null. 

Adequately accounting for spatial heterogeneity and autocorrelation in detection probability is essential for obtaining reliable statistical inference in SCR analyses \citep{moqanaki2021consequences,howe2022estimating}. In the absence of known covariates, the effect of detector-specific variation in detection probability can be modelled by using a function that explains the true pattern of heterogeneity. This function is always unknown and we approximate it using random effects, i.e., by extending SCR with generalized linear mixed models (GLMM). Bayesian implementation of SCR-GLMMs allows modelling and estimation of heterogeneous detection probability surfaces in SCR models  \citep{hooten2003predicting}. Spatially-explicit estimates of detection probability can in turn reveal problematic areas (e.g., regions with very low detection probability), which are important to wildlife monitoring and conservation.

Using simulations, we describe and test three extensions of Bayesian SCR-GLMMs that aim to account for latent spatial heterogeneity in detection probability via the use of random effects: (1) a simple GLMM extension of the basic single-season SCR model by assigning independent random effects (RE) to detector-specific baseline detection probabilities - with the aim to account for unknown spatial variation in detection probability among detectors; (2) a GLMM extension of the basic single-season SCR model incorporating spatial autocorrelation between detectors by means of spatially autocorrelated random effects (SARE), where covariance is modelled as a function of inter-detector distance, thus implicitly defining an ordered neighbourhood structure; (3) a two-group finite mixture (FM) model to identify latent detectability classes of each detector.




We assessed and compared these three structurally different models in terms of (i) their ability to produce unbiased abundance estimates, (ii) their capacity to realistically predict detection probability surfaces, (iii) their model complexity and (iv) their computational overhead. Finally, we considered the role that model comparison could play in selecting the `best' SCR model under different conditions.

\section{Methods}\label{sec:methods}


We first describe a basic single-season SCR model, where we assume a homogeneous baseline detection probability across all the detectors. Following that, we describe three extensions of the SCR model, namely: (i) an SCR-GLMM with independent random effects, (ii) an SCR-GLMM with spatially autocorrelated random effect, and (iii) an SCR-GLMM model with two-group mixture to model detector-specific baseline detection probabilities. Lastly, as a reference point for making comparisons, we outline a special case of the model in (ii), where the known true cause of the variation in detection probability is modelled using fixed effects.



\subsection{Model 1: Basic single-season SCR model (SCR) }\label{sec:snapshotSCR}
A single-season SCR model typically consists of two submodels: a submodel for the spatial distribution of individual ACs within a given habitat $\V \subset \mathbb{R}^2 $, and another submodel for the individual and detector-specific observations, conditional on the location of ACs.

\subsubsection{The ecological submodel}\label{sec:binomialPP}
We considered $N$ individuals to reside in $\V$, each of whom was assumed to move randomly around its AC (with coordinates $\s_i$). Following a homogeneous point process, each individual AC was assumed to be uniformly distributed across the habitat $\V$:
 \begin{align}\label{eq:binomial.pt.proc}
\s_i \sim \text{Uniform} (\V),\, i = 1,2, \dots,N.
\end{align} 
In our analysis, the location $\s_i$ of individual ACs and the number of these ACs ($N$) are both unknown. We used a data augmentation approach to model $N$ \citep{royle2007analysis}, with a large integer $M$ as an upper bound for $N$. We introduced a vector of $M$ latent binary variables $\z = (z_1, z_2, \dots, z_M)'$ such that $z_i = 1$ if individual $i$ is a member of the population and  $z_i=0$ otherwise. Then we assumed that each $z_i$ follows a Bernoulli distribution with inclusion probability $\psi$, the probability that an arbitrary individual from the augmented population of $M$ individuals is a member of the population under study:
 \begin{align}\label{eq:data.augmentation}
\z_i \sim \text{Bernoulli} (\psi).
\end{align} 
Consequently, population size $N = \sum_{i=1}^M z_i$ is a derived parameter, following a binomial distribution with parameters $M$ and $\psi$.

\subsubsection{The observation submodel}\label{sec:obs.submodel}
We considered one sampling occasion and a set of $J$ detectors located in $\V$. The capture history of the $i$-th individual is denoted as $(y_{i1}, y_{i2}, \dots, y_{iJ})$, where each $y_{ij}$ is binary, i.e., $y_{ij}$ is 1 if individual $i$ is detected at detector $j$ and 0 otherwise.
The observed capture-recapture data set, denoted by $\Y_{\text{obs}}$, is of dimension $n\times J$, where $n$ is the number of detected individuals during the SCR survey. We augmented this data set $\Y_{\text{obs}}$ with $M-n$ ``all-zero'' capture histories $\mathbf{0}_J$ following the data augmentation approach. The zero-augmented data set is denoted by $\Y$ and is of dimension $M \times J$. We assumed a Bernoulli model for each $y_{ij}$, conditional on $z_i$:
 \begin{align}\label{eq:bernoulliSCRmodel}
y_{ij} \sim \text{Bernoulli}(p_{ij}z_{i}),
\end{align} 
where $p_{ij}$ denotes the detection probability of the $i$-th individual at the $j$-th detector. 
 The detection probability $p_{ij}$ is a decreasing function of distance, modelled following a half-normal form \citep{efford2004density}:
 \begin{align}\label{eq:halfnormal}
p_{ij} = p_0 \, \exp\Big{(}-\frac{d_{ij}^2}{2\sigma^2}\Big{)}
\end{align} 
where $d_{ij} = d(\s_i, \x_j) = ||\s_i - \x_j||$ is the Euclidean distance between the detector location $\x_j$ and individual AC $\s_i$, $p_0$ is the baseline detection probability, and the scale parameter $\sigma$ quantifies the rate of decline in detection probability $p_{ij}$ with distance $d_{ij}$. The full SCR model can thus be written as:
{\footnotesize
 \begin{align}\label{model:SCR}
    & \psi \sim \text{Uniform} (0,1) \nonumber\\[-0.50em]
    & \sigma \sim \text{Uniform} (0,50) \nonumber\\[-0.50em]
    & \logit(p_0) \sim \N(0, 2) \nonumber\\[-0.50em] 
    & \hspace{-2em} i = 1,2,\dots,M: \nonumber\\[-0.50em]
    & \s_i \sim \text{Uniform} (\V)  \nonumber\\[-0.50em]
    & z_i \sim \text{Bernoulli} (\psi)  \nonumber\\[-0.50em]
    & p_{ij} = p_{0} * \exp(- d_{ij}^2 / (2\sigma^2))  \text{ for } j = 1,2,\dots,J   \nonumber\\[-0.50em]
    & y_{ij} \sim \text{Bernoulli} (p_{ij} * z_i)   \text{ for } j = 1,2,\dots,J  
\end{align} 
}
By modelling detection probability $p_{ij}$ in terms of individual ACs and fitting a decreasing detection function (as in (\ref{eq:halfnormal})) using the distance between ACs and detector location, the SCR model accounts for the spatial autocorrelation within individual capture histories. However, under this model, detection probabilities $p_{ij}$ and $p_{ij'}$ are equal at detectors $j$ and $j'$ whenever the two detectors are located at the same distance from the AC $\s_i$ regardless of other potential sources of variation between the two detectors. In other words, this model does not consider the additional variation in detection probability that may be present at different detectors due to their locations in the landscape and other heterogeneous characteristics.



\subsection{Model 2: Independent random effects SCR model (RE)}\label{sec:RAND}


To account for spatial heterogeneity in detection probability, causing detector-specific variation in detection probabilities, we used a simple GLMM extension of the basic single-season SCR model (Model 1). Here, we assigned a logistic-regression type model to baseline detection probability for each detector:
 \begin{align}\label{model:RE}
& \logit(p_{0j}) = \mu + W_j,\, j = 1,2,\dots,J 
\end{align} 
where $\mu$ denotes the intercept and $W_j$ denotes the random effect for the $j$-th detector. The detection probability $p_{ij}$ for individual $i$ at detector $j$ is expressed as 
 \begin{align}\label{eq:halfnormal.RE}
p_{ij} = p_{0j} \, \exp\Big{(}-\frac{d_{ij}^2}{2\sigma^2}\Big{)}.
\end{align} 
We assumed a $\N(0, \sigma_w^2)$ prior for each $W_j$, $j = 1,2,\dots,J$ and a $\N(0, 2^2)$ prior for $\mu$. The variance parameter $\sigma_w^2$ can be given a weakly informative prior.  We referred to this model as independent random effects SCR model (RE). Note that, RE model does not specifically account for spatial autocorrelation in detection probability across detectors.


%
%

\subsection{Model 3: Spatially autocorrelated random effects SCR model (SARE)}\label{sec:SARE}

\begin{figure}[H] 
	\centering
	{\footnotesize \begin{tabular}{@{\hspace{-0.1cm}} l } 
	\includegraphics[scale=0.4]{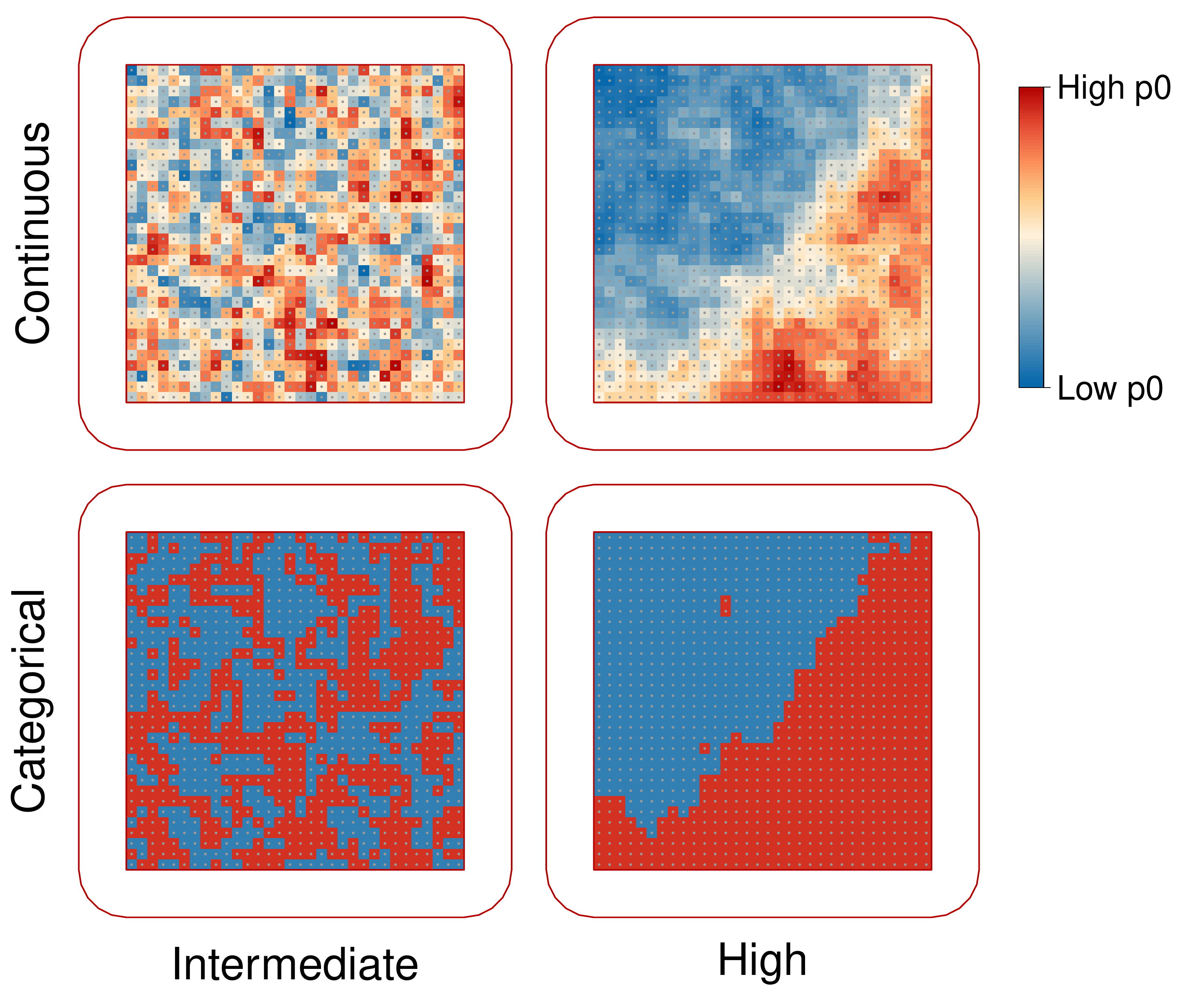}
		\end{tabular}}
	\caption{\footnotesize Examples of spatially variable and autocorrelated
baseline detection probability surface: $\mathbf{p_0} = (p_{01}, p_{02}, \dots, p_{0J})'$. The color gradient corresponds to different values of baseline detection probability. The surface is overlayed on a grid of detectors (gray dots) centered in a habitat (entire area surrounded by the red line with rounded corners). Shown in rows, spatial variation in detection probability can be either continuous or categorical (with 50\% of the detectors remaining inactive while the rest have a constant baseline detection probability). Shown in columns, spatial autocorrelation may vary from intermediate (Moran’s I $\approx$ 0.3) to high (Moran’s I $\approx$ 1).}
\label{fig:sample.autocorrelated.det}

\end{figure}

\begin{figure}[H] 
	\centering
\hspace{-0.98cm}
\begin{minipage}[c]{0.75\textwidth}
\includegraphics[scale=0.25]{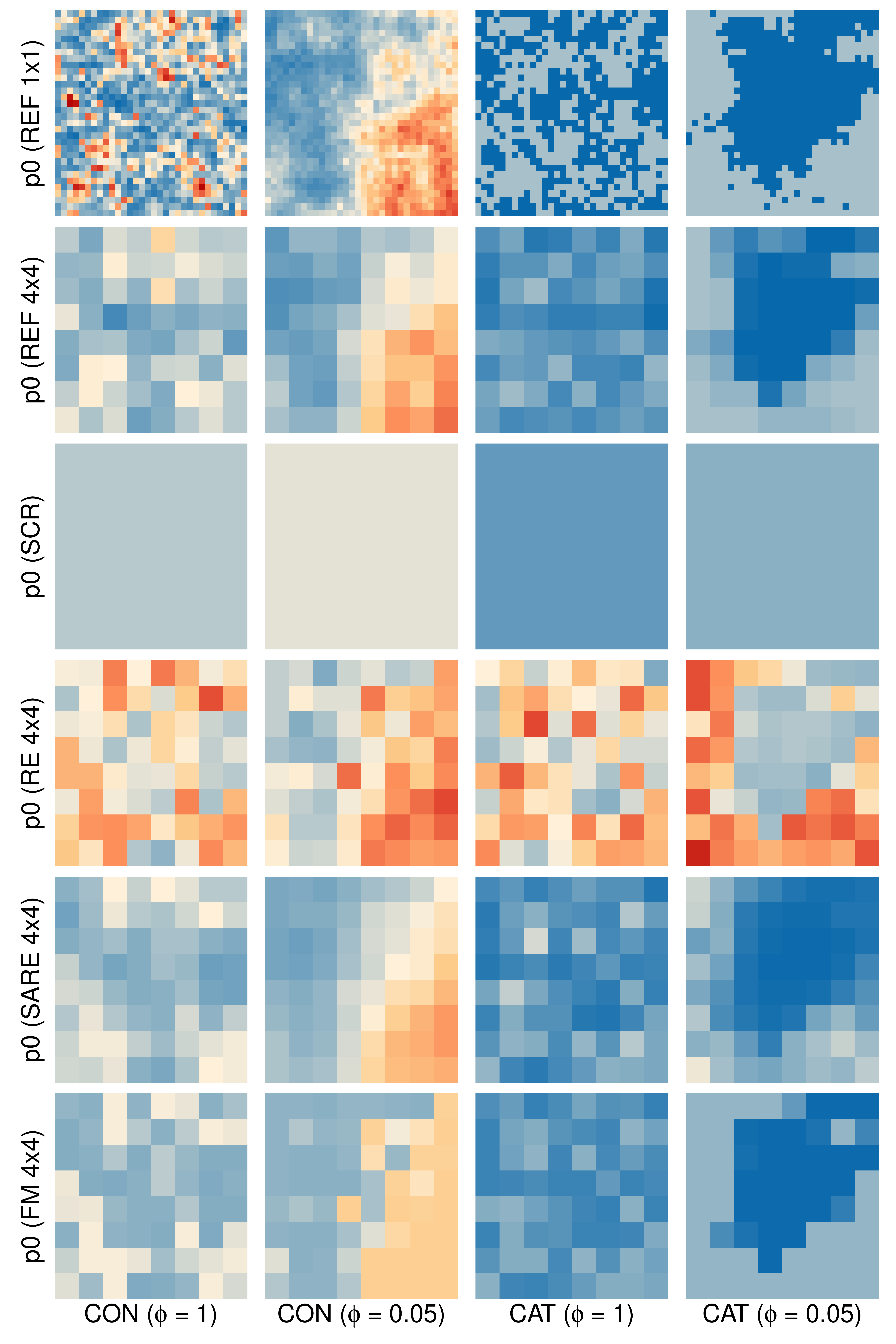} 
\end{minipage} 	\hspace*{1.4cm}
\begin{minipage}[c]{0.2\textwidth}
\vspace*{-15.5cm}	\includegraphics[scale=0.75]{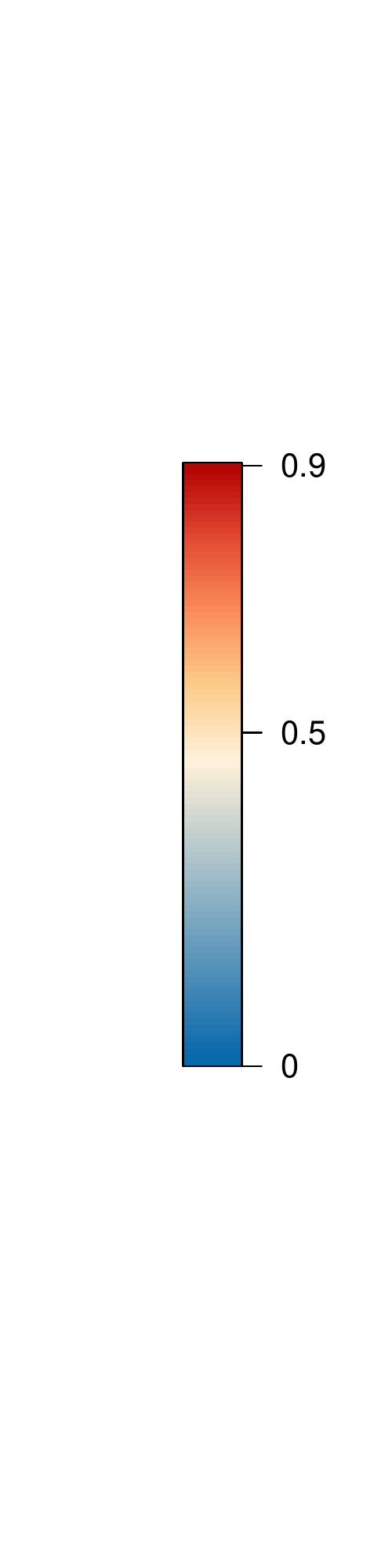} 
\end{minipage} 	
	\caption{\footnotesize Illustration of baseline detection probability surfaces for simulation scenarios under average baseline detection probability $\eta = 0.3$. In rows: simulated baseline detection probability surface (`REF $1\times1$'), baseline detection probability surface after averaging the simulated values for each cluster of detectors at $4\times4$ scale (`REF $4\times4$'), predicted baseline detection probability surface from four models: SCR, RE (aggregation $4\times4$), SARE (aggregation $4\times4$), FM (aggregation $4\times4$). Labels on the $x$-axis refer to scenarios with continuous (``CON'') and categorical (``CAT'') detector-specific variation in detection probability. Second and fourth columns represent high autocorrelation among detectors (i.e., under $\phi = 0.05$), whereas first and the third column represent intermediate autocorrelation (i.e., under $\phi = 1$). Colors correspond to different values of baseline detection probability.}

\label{fig:predictedp0Surfaces.p0.3}
\end{figure}

We extended the basic single-season SCR model (Model 1) to account for spatial autocorrelation among detectors. In particular, we developed an SCR model for situations, where detectors at close proximity are more likely to have similar detection probability as compared to more distant detectors. We modelled this spatial autocorrelation by introducing an autocorrelated random effect $\W = (W_1, W_2, \dots, W_J)'$ of length $J$. We assumed $W$ to follow a multivariate normal distribution with mean $\mathbf{0}_J$ and covariance matrix $\Gamma = ((\gamma_{jk}))$, which controls the spatial dependence between detectors. We modelled each element $\gamma_{jk}$ of this covariance matrix as a decreasing function of distance between detectors $j$ and $j'$ following \cite{moqanaki2021consequences},
 \begin{align}\label{eq:covariance.function}
\gamma_{jj'} = \exp (- \phi \, \delta_{jj'} )
\end{align} 
where $\delta_{jj'} = d(\x_j, \x_{j'}) = ||\x_j - \x_{j'}||$ is the Euclidean distance between the detector locations $\x_j$ and $\x_{j'}$. This covariance function implicitly defines an ordered neighbourhood for each detector and $\phi$ controls the rate of distance-dependent decay of spatial autocorrelation between the detectors. In particular, detectors are highly autocorrelated if $\phi$ is small (e.g., 0.05), and autocorrelation decreases as $\phi$ increases (Figures~\ref{fig:sample.autocorrelated.det} and \ref{fig:predictedp0Surfaces.p0.3}).  Similar to the RE model, the detection probability $p_{ij}$ for individual $i$ at detector $j$ is then expressed as
 \begin{align}\label{eq:halfnormal.SARE}
p_{ij} = p_{0j} \, \exp\Big{(}-\frac{d_{ij}^2}{2\sigma^2}\Big{)},
\end{align} 
where 
 \begin{align}\label{eq:baselineDetProb.SARE}
\logit(p_{0j}) = \mu + W_j,\, j = 1,2,\dots,J.
\end{align} 

Here, we assigned a $\N(0, 2^2)$ prior for $\mu$ and a $\N(0, 5^2)$ prior for log-transformed $\phi$.
We referred to this model as spatially autocorrelated random effect SCR model (SARE).
%
%


When $\phi = 0$, each component $\gamma_{jj'}=1$ (for any $j$ and $j'$), the random effect $\W$ becomes a degenerate process, implying exact dependence between the detectors. Hence, the value of each random effect $W_j$ is identical at any location of the detector grid. This is equivalent to basic single-season SCR model (Model 1), where we use a homogeneous baseline detection probability $p_0$ for each detector in the detector grid. Conversely, when $\phi \rightarrow \infty$, covariance matrix $\Gamma$ reduces to an identity matrix, and consequently, SARE model reduces to GLMM with independent random effects.







\subsection{Model 4: Two-group finite mixture SCR model (FM)}\label{sec:FM}
Variable sampling intensity could be associated with ordered classes of unknown variation in detection probability across the landscape.
For our study, we proposed using a two-group finite mixture SCR model (FM) to model heterogeneity in detection probability between detectors \citep{cubaynes2010importance, turek2021bayesian}. Here, we defined two groups of heterogeneity, viz., 1 and 2 assuming first group to have lower detection probability than the second one. We introduced two detection probability parameters $\eta_1$ and $\eta_2$, where $\eta_k$ is the detection probability of the $b$-th subgroup, $b = 1,2$. A constraint is imposed on these parameters $\eta_1 \leq \eta_2$ to ensure identifiability. Further, we defined binary indicator variables $u_j$ ($j=1,2,\dots,J$) to indicate the subgroup that a detector belongs to:
 \begin{align}\label{model:FM}
p_{0j} = (1-u_j)\, \eta_1 + u_j \, \eta_2, \, j = 1,2,\dots,J.
\end{align} 
MCMC computation allows the binary classification in our two-group mixture model to implicitly account for the group membership probabilities $\pr(u_j=1)$ and consequently, allows estimation of each $p_{0j}$ via (\ref{model:FM}) accounting for the uncertainty in the group membership probabilities of each detector. This provides a flexible approach of estimating spatial heterogeneity in detection probability surface. We assigned a Bernoulli prior to each $u_j$ with probability $\pi$ of being assigned to second group. Further, we assumed weakly informative bounded uniform priors for the probability parameters $\eta_1, \eta_2$ and $\pi$.

\subsection{Model 5: SCR model with known true effects (FE)} \label{sec:FE}

For the sake of assessing and comparing the performance of the above models, we considered a GLM extension of basic single-season SCR model (Model 1) using  detector-specific effects (the true source of variation) to model baseline detection probability. This can be executed by supplying the known simulated effect $\W$ as an observed ``virtual" covariate and then model the baseline detection probability: $\logit(p_{0j}) = \mu + W_j$. Consequently, the detection probability $p_{ij}$ is expressed as: $p_{ij} = p_{0j} \, \exp(-d_{ij}^2/(2\sigma^2))$.
%
%
The rest of the model remains the same as Model 1 and we referred to this model as FE.


\section{Simulation study}\label{sec:simstudy}
For simulations, we used a $32 \times 32$ detector array (number of traps $J$ = 1024) with 1 distance unit (du) of minimum inter-detector spacing. The detector array is centred on a $41 \times 41$ du habitat, surrounded by a 5-du habitat buffer (Figure~\ref{fig:sample.autocorrelated.det}). We used a $\sigma$ value of 1.5 for all the simulations so that the buffer width is larger than $3\sigma$ resulting in negligible detection probability of individuals with AC near the habitat boundary \citep{efford2011estimation}. We simulated SCR data sets for $N=300$ individuals leading to a population-level home range overlap index $k=\sigma \sqrt{\text{Density}} = 0.63$ \citep{efford2016density}. We set the size of the augmented population $M$ to be 500.



\subsection{Simulation scenarios}\label{sec:simscenarios}
For each simulation, we used the SARE model (Model 3, Section~\ref{sec:SARE}) to generate SCR data with spatially autocorrelated detection probability between detectors. We created simulation scenarios by varying spatial autocorrelation rate parameter $\phi$ with high ($\phi = 0.05$) and intermediate ($\phi = 1$) spatial autocorrelation to simulate spatially varying random effect $\W = (W_1, W_2, \dots,W_J)'$ (Figure~\ref{fig:sample.autocorrelated.det}). 

\subsubsection{Continuous detector-specific variation in detection probability}\label{sec:simscenarios.cont}

Detection probability may exhibit continuous spatial variation if it is linked with underlying habitat characteristics, such as elevation, forest cover, or distance from roads that influence animal behavior or detection effort and efficiency \citep{moqanaki2021consequences}. For simplifying the interpretation of $\mu$ in SARE (Model 3), we transformed it into a new variable $\eta$ via the link $\mu = \logit(\eta)$. Here, $\eta$ can also be viewed as the average baseline detection probability, providing a clearer interpretation for the readers. In simulations, we used three values of $\eta$ to generate low ($\eta = 0.1$), intermediate ($\eta = 0.3$), and high ($\eta = 0.6$) baseline detection probability for each detector, subject to spatial autocorrelation infused by $\W$ (Figure~\ref{fig:sample.autocorrelated.det}, row 1). 


\subsubsection{Categorical detector-specific variation in detection probability}\label{sec:simscenarios.cat}

Discrete differences in sampling or environmental characteristics can lead to categorical classes of variation in detection probability between detectors. We considered an extreme case, where 50\% of the detectors would remain inactive, and the remaining detectors would have a constant detection probability \citep{moqanaki2021consequences}. Thus, a portion of the study area would remain entirely unsampled. For simulating such scenarios, we transformed each $p_{0j}$ into a discrete variable taking only one of the two values 0 and $\logit(\eta)$ to create two classes of detector-specific baseline detection probability using (\ref{eq:baselineDetProb.SARE}):
 \begin{align}
p_{0j} = 
\begin{cases}
    0, & \text{ if } W_j \leq q_{50} \nonumber \\
    \logit(\eta), & \text{otherwise}
\end{cases}
\end{align} 
where $q_{50}$ is 50\% quantile of the effect $W_j$'s.
We used two values of $\eta$ to generate low ($\eta = 0.1$) and intermediate ($\eta = 0.3$) level of baseline detection probability for each detector (Figure~\ref{fig:sample.autocorrelated.det}, row 2).

In summary, we divided all the simulation scenarios in two broad setups, viz. continuous and categorical, with respect to detector-specific variation in detection probability. In the continuous setup (`CON'), we generated six simulation scenarios by combining two levels of autocorrelation $\phi$ and three levels of detection $\eta$. In the categorical setup (`CAT'), we generated four simulation scenarios by combining two levels of $\phi$ and two levels of $\eta$. Thus, in total, we generated 10 simulation scenarios. For each simulation scenario, we generated 100 independent SCR data sets, resulting in 1000 simulated SCR data sets in total (Table~\ref{table:ndets}).

\subsection{Curse of dimensionality}\label{sec:scaling}

In many SCR studies the majority of detectors are associated with no or very few detections \citep{gerber2015spatial,tourani2022review}. In such situations, fitting complex models such as SARE, FM and RE, which involves large number of parameters and latent variables, may lead to poor Markov chain Monte Carlo (MCMC) convergence, below par mixing, and over-fitting. This phenomenon is known as the curse of dimensionality and expected to occur when models are over-parameterised \citep{wikle2010general}.

We mitigated this issue by dimension reduction of the random effects. To do this, we aggregated random effects that are used to model baseline detection probability, such that a single random effect value is assigned to a cluster of neighboring detectors. Note that we are not aggregating detections themselves. For instance, in SARE, if each cluster contains $n_c$ detectors, then each detector belonging to a cluster (say, $j$-th) will share the same random effect value $W_j$, $j = 1,2, \dots, J/n_c$ ($J$ being the total number of detectors).
Here, we aggregated the random effects by a factor of 4 (squares of $4\times4$ detectors = one cluster). When aggregated, the $32\times32$ detector  grid  (i.e., 1024 detectors, Figure~\ref{fig:sample.autocorrelated.det}) forms a grid of $8\times8$ clusters. 




\subsection{Model fitting description}\label{sec:model.fitting}
We fitted five SCR models to the same simulated datasets: (i) basic single-season SCR model without aggregation (Section~\ref{sec:snapshotSCR}), (ii) RE model, both with and without aggregation (Section~\ref{sec:RAND}), (iii) SARE model, both with and without aggregation (Section~\ref{sec:SARE}), (iv) FM model, both with and without aggregation, and (v) FE model without aggregation (Section~\ref{sec:FE}). The models were fitted using MCMC simulation with NIMBLE \citep{valpine2017nimble, nimbleSoftware2021} in R version 3.6.2 \citep{Rsoftware2019}. We used the R package nimbleSCR \citep{nimblescr2020bischof, turek2021bayesian}, which implements the local evaluation approach \citep{milleret2019local} to increase MCMC efficiency. For each simulated data set, we ran three chains of (i) 30,000 iterations for both basic single-season SCR and FE model including burn-in 12,000 iterations, (ii) 100,000 iterations for SARE and RE including burn-in 20,000 iterations (both with and without aggregation), (iii) 60,000 iterations for FM (without aggregation) including burn-in 12,000 iterations and 20,000 iterations for FM (with aggregation) including 4,000 iteration burn-in.
MCMC convergence of each model was monitored using the Gelman-Rubin convergence diagnostics $\hat{R}$ \citep[with upper threshold 1.1,][]{gelman2014bayesian} and visual inspection of traceplots.




\subsection{Model performance measures}\label{sec:model.performance.measures}
We used relative bias, coefficient of variation, and coverage probability to evaluate the performance of each fitted models with respect to estimation of focal parameters (e.g., population size, $\sigma$). Suppose $\{\theta^{(r)} \, : \, r = 1, 2, \dots, R\}$ denotes a set of MCMC draws from the posterior distribution of a scalar parameter $\theta$. 

\tbf{Relative bias}.
	Relative bias (RB) is calculated as 
	 \begin{align}
	\widehat{\text{RB}} (\theta) = \frac{\hat{\theta} - \theta_0}{\theta_0},
	\end{align} 
	where $\hat{\theta}$ denotes the posterior mean $\frac{1}{R} \sum_{r=1}^{R} \theta^{(r)}$ and $\theta_0$ gives the true value.

\tbf{Coefficient of variation}.
	Precision was measured by the coefficient of variation (CV):
		 \begin{align}
	\widehat{\text{CV}} (\theta) = \frac{\widehat{\text{SD}}(\theta)}{\hat{\theta}},
	\end{align} 
	where $\widehat{\text{SD}}(\theta) = \sqrt{\frac{1}{R} \sum_{r=1}^{R-1} (\theta^{(r)} - \hat{\theta})^2}$ is the posterior standard deviation of parameter $\theta$.

\tbf{Coverage probability}.
Coverage probability was computed as the proportion of converged model fits for which the estimated 95\% credible interval (CI) of the parameter $\theta$ contained the true value of $\theta_0$.

\subsubsection{Effective sample size and MCMC efficiency}\label{sec:ess}

 To compare the efficiency of the different MCMC algorithms of the fitted models, we computed the effective sample size (ESS) and MCMC efficiency (= ESS/MCMC run time) of each top-level parameter for each of the model runs. We used the `effectiveSize' function from the R package coda to compute ESS \citep{coda2006}. The calculation of ESS is based on the combined samples of the converged MCMC chains after discarding the burn-in period. The MCMC computation time was calculated excluding the burn-in period. To obtain stable estimates of the quantities of interest, it is recommended to have ESS greater than 400 \citep{vehtari2021newRhat}.

Although MCMC algorithms are used to generate samples from posterior distributions, efficiency can vary between different algorithms. There are two primary measures of efficiency of MCMC algorithms - quality of MCMC mixing and speed of MCMC computation. We computed `MCMC efficiency' as a combined metric to assess both of these characteristics of MCMC algorithms, so that we could compare the efficiency of different MCMC algorithms. For our study, we reported the mean MCMC efficiency for each top-level parameters in the model and across all the converged replicates in each scenario.  



\subsubsection{Spatial accuracy of predicted baseline detection probability surfaces}\label{sec:method.compare.det.maps}

Baseline detection probability surfaces obtained from SCR analyses are useful in evaluating the performance of our SCR models as they have the potential to reveal spatial patterns in detection probability (such as pockets with very low/high detection probability) that could be of practical relevance. We compared the accuracy of the detector-specific baseline detection probability surfaces predicted by the different models with the true simulated surface $(p_{01},p_{02},\dots,p_{0J})'$. We quantified the accuracy by calculating the expected sum of squared errors (SSE). In practice, we first obtained posterior MCMC samples of baseline detection probability surface $(p_{01}, p_{02},\dots, p_{0J})'$ and compute mean squared error for detector $j$: $\text{SSE}_j = \frac{1}{R} \sum_{r=1}^{R} (p_{0j}^{(r)} - p_{0j})^2$, where $\{p_{0j}^{(1)}, p_{0j}^{(2)}, \dots, p_{0j}^{(R)} \}$ denotes posterior MCMC sample of $p_{0j}$, $j=1,2,\dots,J$. Finally, we calculated total error sum of squares $\text{SSE} = \sum_{j=1}^{J} \text{SSE}_j$ as a measure of predictive accuracy of detection probability surface. Smaller SSE implies a more accurate prediction (closer to the truth) of the baseline detection probability surface. We used $\Delta \text{SSE}$, relative to the model with the lowest SSE ($\Delta \text{SSE}$ $= \text{SSE} - \min\{\text{SSE}\}$), to compare the accuracy of predicted baseline detection probability surface amongst the different models.

 \subsubsection{Model comparison using WAIC}\label{sec:model.comparison.waic}


We compared the fitted models using Watanabe-Akaike information criterion (WAIC) \citep{watanabe2010asymptotic}, which is computed as
 \begin{align}\label{waic} 
 \text{WAIC} = -2\sum_{i=1}^{M} \log \big{(}  \frac{1}{R} \sum_{r=1}^{R} f(\mathbf{Y}_i \, | \, \boldsymbol{\ttheta}^{(r)}) \big{)} + 2 \, p_w.
\end{align} 
where $f(\mathbf{Y}_i \, | \, \boldsymbol{\ttheta})$ denotes the likelihood of $i$-th individual capture history $\mathbf{Y}_i = (y_{i1}, y_{i2}, \dots, y_{iJ})'$ in the model. Here, we adopt the second of the two variants of the penalty term $p_w$ proposed by \cite{gelman2014understanding}:\vspace*{-0.1em}
 \begin{align}\label{p.waic} 
p_w = \sum_{i=1}^{M} \Big{\{}\frac{1}{R-1} \sum_{r=1}^{R}\Big{(}\log f(\mathbf{Y}_i \, | \, \boldsymbol{\ttheta}^{(r)}) - \frac{1}{R} \sum_{r=1}^{R} \log f(\mathbf{Y}_i \, | \, \boldsymbol{\ttheta}^{(r)}) \Big{)}^2\Big{\}}
\end{align} 
A model with smaller WAIC is preferred. We use $\Delta \text{WAIC}$ ($= \text{WAIC} - \min\{\text{WAIC}\}$) to compare the different models in terms of their model fit and complexity.




\section{Results}\label{sec:results}


During comparison and interpretation, we only considered models that had reached convergence and exhibited proper mixing of all the top level parameters (e.g., $N$, $\sigma$, $\phi$, $\eta$), with $\hat{R} \leq 1.1$. While all SCR (Model 1) and FE models (Model 5) converged, convergence of the models SARE, RE and FM were found to be challenging without aggregating the random effects (Table~\ref{table:nconverged}). Only under the extreme categorical scenarios with low baseline detection probability ($\eta=0.1$), the convergence rates (i.e., the number of converged models out of 100 repetitions) of RE (Model 3) were found to be higher ($\geq60\%$) than the other two GLMM models. The convergence rate of RE and FM improved substantially when random effects were aggregated (66 -- 100\%). Convergence rate for the SARE improved substantially  (67 -- 96\%) after aggregating the random effects under high spatial autocorrelation scenarios ($\phi = 0.05$), whereas the improvement was less pronounced (5 -- 39\%) under intermediate autocorrelation scenarios ($\phi = 1$).   

For all the models that converged, mean ESS were considerably higher than the suggested threshold of 400, indicating that the MCMC chains were long enough to provide stable estimates (Tables~S6 -- S9, Supp. material). SCR and FE had the highest MCMC efficiency (mean MCMC eff. $>0.8)$ under most scenarios except for the extreme categorical scenario with $\eta=0.1$ and $\phi = 0.05$, where SCR had a relatively lower mean MCMC efficiency 0.41. MCMC efficiency of both SARE and RE models were 1.5 -- 2 times lower than SCR and FE models in these scenarios despite aggregating of the random effects. Among the three GLMM formulations, SARE showed the highest MCMC efficiency in most scenarios (mean 0.8 -- 1.6 in continuous and 0.27 -- 0.77 in categorical scenarios). FM model had the lowest MCMC efficiency (mean 0.01 -- 0.1) across all scenarios, primarily due to the higher MCMC computation time (Table S9). Considering the overall poor MCMC convergence of the three GLMMs when fitted without aggregation, we only considered the results with dimension reduction.

\subsection{Estimates of population size}\label{sec:par.estimates}
\begin{figure}[H] 
	\centering
	{\footnotesize \begin{tabular}{@{\hspace{-0.1cm}} l } 
 	\includegraphics[scale=0.8]{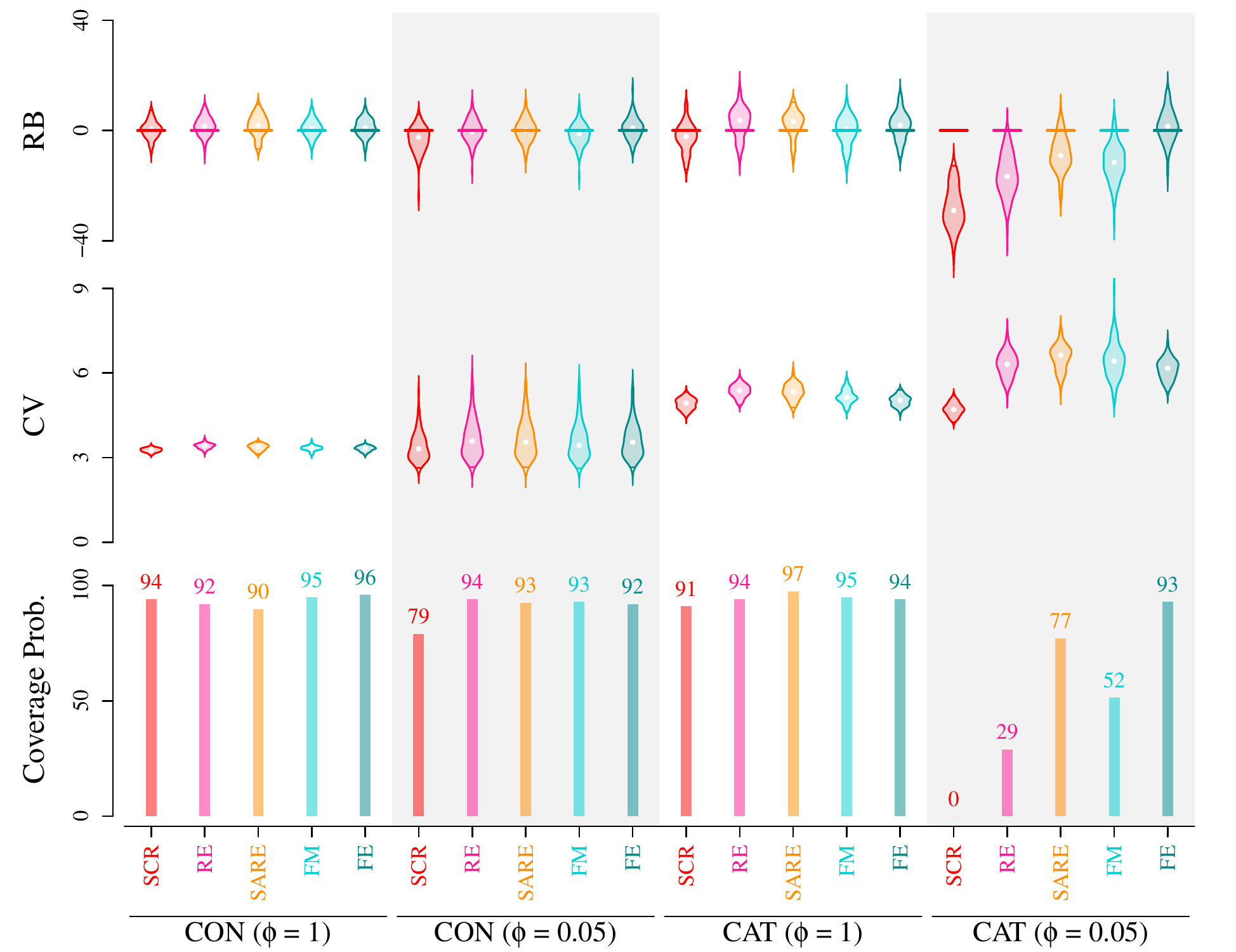}
		\end{tabular}}
	\caption{\footnotesize Posterior summaries of population size $N$ derived from five models (i) SCR, (ii) RE (aggregation $4\times4$), (iii) SARE (aggregation $4\times4$), (iv) FM (aggregation $4\times4$), (v) FE. From top to bottom, results compare relative (1) bias (RB, in \%), (2) coefficient of variation (CV, in \%), (3) coverage probability (in \%) of 95\% credible interval (CI) for simulation scenarios under average baseline detection probability $\eta = 0.3$. Violins represent the distribution of RB and CV from 100 simulations. Labels on the $x$-axis refer to scenarios with continuous (``CON'') and categorical (``CAT'') detector-specific variation in detection probability. Grey shaded background indicates scenarios with high autocorrelation among detectors ($\phi = 0.05$), whereas white background indicates scenarios with intermediate autocorrelation ($\phi = 1$). All results shown are based on models that met convergence criteria.}
\label{fig:Nest.p0.3}
\end{figure}

All five models showed negligible bias in population size $N$ estimates under most simulation scenarios tested here. Both SARE and FE (models that distinctly account for spatial autocorrelation) estimated the population size with moderate accuracy across all the scenarios (median RB: -$9\%$ -- $6\%$) (Tables~S3, S5). Although SCR and RE did not specifically model spatial autocorrelation between detectors, population size estimates from these models showed negligible bias (median RB: -$10$ -- $5\%$) in most scenarios considered (Tables~S1 and S2, Supp. material). 
However, under scenarios with categorical spatial variation in baseline detection probability and high autocorrelation ($\phi=0.05$), SCR model showed approximately 30\% negative bias in estimating population size (Figure~\ref{fig:Nest.p0.3}). RE model produced an elevated negative bias (median RB: -17\%) under the categorical scenario with $\eta=0.3$ and $\phi=0.05$ (Table~S2). FM also showed a similar level of accuracy in estimating population size compared to the SARE and FE models for all the continuous scenarios (Table~S4). Although FM seemed to be structurally better suited for the scenarios with categorical variation in detection probability (due to the integration of membership in discrete detectability groups), it showed an 11\% negative bias in each of the categorical scenarios with high autocorrelation. 


The coefficient of variation (CV) in population size estimated with the five fitted models varied moderately (median CV: 3 -- 16\%) under $\eta = 0.1$ and was less than 8\% for the remaining scenarios with $\eta \geq 0.3$. Coverage probabilities for SCR were $> 90\%$ for the scenarios with intermediate autocorrelation ($\phi= 1$). But when spatial autocorrelation was high ($\phi= 0.05$), coverage declined drastically (65 -- 97\% coverage) for the continuous scenarios and dropped to less than $20\%$ for the categorical scenarios (Table~S1). Coverage probabilities for SARE, RE and FE were $\geq 90\%$ (coverage for FM $\geq 81\%$) for all the scenarios except for the extreme categorical scenario with $\eta=0.3$ and $\phi = 0.05$, where coverage probabilities for SARE, RE and FM were 0.77, 0.29 and 0.51, respectively (Tables~S2-S5).

\subsection{Detection probability surfaces and model comparison}\label{sec:results.compare.det.maps}

\begin{figure}[H] 
	\centering
	{\footnotesize \begin{tabular}{@{\hspace{-0.1cm}} l } 
 	\includegraphics[scale=0.8]{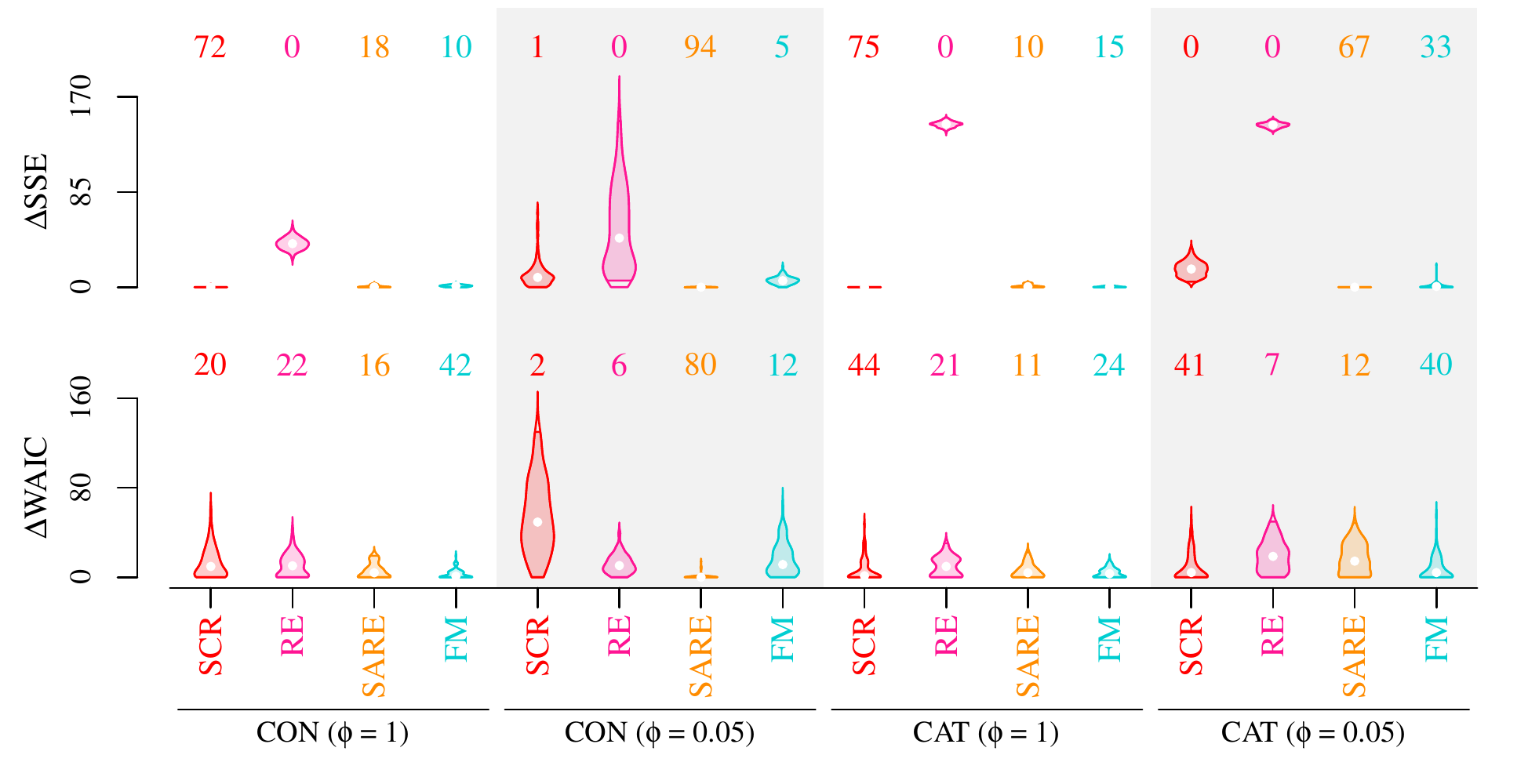}
		\end{tabular}}
	\caption{\footnotesize  $\Delta\text{SSE}$ and $\Delta\text{WAIC}$ from four models (i) SCR, (ii) RE (aggregation $4\times4$), (iii) SARE (aggregation $4\times4$), (iv) FM (aggregation $4\times4$) for simulation scenarios under average baseline detection probability $\eta = 0.3$. Violins represent the distribution over 100 replicated data sets in each scenario. Labels on the $x$-axis refer to scenarios with continuous (``CON'') and categorical (``CAT'') detector-specific variation in detection probability. Grey shaded background indicates scenarios with high autocorrelation among detectors ($\phi = 0.05$), whereas white background indicates scenarios with intermediate autocorrelation ($\phi = 1$). All results shown based on models that met convergence criteria.}
\label{fig:SSE.WAIC.p0.3}
\end{figure}

Both SARE and FM models produced reliable detection probability surfaces in the presence of high spatial autocorrelation between detectors. SARE-generated surfaces were more accurate in estimating surfaces of baseline detection probability, with the lowest SSE in 65 -- 94\% of the replicates in both continuous and categorical scenarios with high autocorrelation. Although FM were more precise than SARE and RE in scenarios with intermediate autocorrelation, SCR (which assumes homogeneous baseline detection probability) had the lowest SSE in 72 -- 97\% of the replicates in these scenarios (Figure~\ref{fig:SSE.WAIC.p0.3}, S2). 

Under the scenarios with continuous spatial variation in detectability and high autocorrelation, SARE was selected 4 -- 6 times more frequently than the other models in model comparison based on WAIC when $\eta \geq 0.3$. With intermediate autocorrelation, FM and RE were selected 1.5 -- 2 times more frequently than the other models when $\eta$ was 0.3 and 0.6, respectively. For all remaining scenarios (including the scenarios with categorical spatial variation), SCR (Model 1) was selected by WAIC. 





\section{Discussion}\label{sec:discussion}

Using a simulation study, we developed and tested three SCR-GLMMs (RE, SARE and FM) to assess their performance in accounting for latent spatial heterogeneity and autocorrelation in detection probability among detectors. SARE (Model 3), the data generating model, was the most reliable model in estimating population size across all the tested scenarios.  When  autocorrelation was high  ($\phi = 0.05$), SARE also performed best in predicting the baseline detection probability surface (as indicated by SSE). The population size estimates from RE and FM (Models 2 and 4) were largely unbiased in the presence of continuous detector-specific variation in baseline detection probability surface, but the estimates were subject to a pronounced negative bias when fitted under the extreme scenarios with categorical variation and high autocorrelation. FM outperformed  SARE and RE in terms of SSE for predicted surfaces of baseline detection probability when autocorrelation was at intermediate level ($\phi = 1$).

Unknown latent and autocorrelated variation in detection probability among detectors in an SCR survey is common \citep{gaspard2019residual}, and can remain present in SCR data due to sampling design or landscape characteristics. As shown by \cite{moqanaki2021consequences}, failure to properly account for spatially autocorrelated detection probability may result in biased and overdispersed population size estimates (Figure~\ref{fig:Nest.p0.3}). In this study, we presented a Bayesian SCR-GLMM (viz., SARE, Model 3) that specifically accounts for the spatial autocorrelation between detectors. The primary advantages of modelling spatial autocorrelation among detectors include the ability to use information on detector configuration to correctly account for uncertainty in the estimates. In a practical context, this may aid identification of locations or regions inside the study area without any detection record. Fitting SCR (Model 1) in cases of high autocorrelation produced a 30\% RB with approximately zero coverage probability, whereas the SARE (Model 3) showed less than 10\% RB and greater than 77\% coverage probability. Even models that allow variation among detectors but do not explicitly account for spatial autocorrelation (RE and FM) were able to produce estimates of population size with little bias for the majority of the simulated data sets, thus showcasing the potential of SCR-GLMMs models.

In large-scale monitoring programs, data often hail from both structured and unstructured or opportunistic sampling \citep{altwegg2019occupancy, bischof2020estimating, isaac2020data}. In certain extreme cases (e.g., citizen science data), large portions of the study area may be left unsampled, unbeknownst to the investigator \citep{johnston2022outstanding, bird2014statistical}. The three SCR-GLMMs tested here (SARE, RE and FM) allow modelling unknown spatial variation in detection probability in the absence of known fixed effects. SCR-GLMMs can also help quantify this unknown spatial variation in detectability. Spatially-explicit estimates of detection probability obtained with SCR-GLMMs can be useful in planning and adjusting large-scale surveys, if they help investigators identify regions with high and low detection probability, including apparent holes in sampling. On the flip side, Bayesian SCR-GLMMs involve a large number of unknown parameters, making these models challenging to fit, manifested in slow computation speeds and convergence issues in certain conditions (e.g., SCR data with low number of detections per detector, fitting of GLMM models without aggregating the random effects). For choosing a model, practitioners will need to weigh the benefits of accounting for spatial heterogeneity in detection probability against the costs associated with model complexity.

 High dimensional random effects models can easily overfit typical SCR data with few or no detections at the majority of detectors. Dimension reduction of the random effects is a typical strategy to avoid overfitting and to control the number of random effects in a model (Section~\ref{sec:scaling}) \citep{hefley2017basis, gelman2014bayesian}. Pooling information allows reliable inference from model fitting that would otherwise be computationally unstable as shown from the improved convergence rates of all SCR-GLMM models when random effects were aggregated (Table~\ref{table:nconverged}). The choice of aggregation level implies a trade-off between sample size per detector (high aggregation to achieve dimension reduction) and the resolution of spatially explicit estimates of detectability (low aggregation for more spatial detail). We recommend increasing the aggregation level until the MCMC convergence criteria are met for the key parameters of interest. This way the aggregation level is kept as small as possible so that model fitting is possible and the level of detail meet the requirements of the investigation. In empirical analyses, random effects can also be aggregated based on natural groups of detectors, such as administrative units, sub-regions that differ in varying sampling effort or some other categorical factors.

For the SARE model, we advise caution in choosing an upper bound for the aggregation scale as the spatial autocorrelation is specifically modelled as a decreasing function of inter-detectors distance. It may get computationally intractable to estimate model parameters with low number of random effects since the fitted coarse surface would over-dilute the true scale of variation in the autocorrelated surface. Further, strong negative bias may arise in the estimate of population size under highly autocorrelated scenarios, similar to what we experienced when fitting the basic SCR model to our simulated data with heterogeneous and spatially autocorrelated detection probability.

When the number of detections per detector in SCR data sets is low, multicollinearity can occur between the detector- or cluster-specific random effects and other parameters in half-normal detection function. For instance, such multicollinearity arise in situations where SARE is fitted to SCR data sets that are not sufficiently informative to reveal underlying autocorrelation amongst detectors. Based on SSE values and WAIC in our study, we recommend fitting SARE models primarily to data from extreme sampling situations where both detection probability and spatial autocorrelation are high. In all other situations, SARE is expected to give a poorer fit (and poor MCMC convergence), whereas basic single-season SCR can cope with moderate levels of variation present among the detectors even under low detectability (as indicated by SSE and WAIC; Figure~S3, Supp. material). Overall, we found WAIC to be useful in selecting the best model in scenarios with different levels of autocorrelation, which holds promise for WAIC application in empirical analyses. 



In this study, we focused on three extensions of the SCR model that can account for latent heterogeneous detection probability. Other potential modelling solutions for dealing with a lack of covariates include: a. Bayesian nonparametric models allowing for the possibility of infinite number of subgroups for the detection probability \citep{turek2021bayesian}, b. conditionally autoregressive random effects model (CAR) that specifically models spatial autocorrelation between detectors \citep{nicolau2020incorporating}, and c. basis function models by obtaining basis expansion from factorization of a pre-specified correlation matrix \citep{hefley2017basis}. Recently, \cite{stevenson2021spatial} developed an SCR model that models spatially autocorrelated detections based on Gaussian random fields. While the modelling approach can be advantageous in situations where variation in detection probability occurs regularly within individual home ranges, the latent detection field SCR model requires integrating out the spatially autocorrelated random effects as well as the ACs, resulting in a significant computational burden. Each of these different classes of models is computationally extensive, overparamterized and likely to overfit the sparse SCR data sets that are common in ecological studies \citep{gerber2015spatial,tourani2022review}, but we anticipate future advancements can overcome the computational and/or modelling barriers to facilitate successful application of these sophisticated techniques to model heterogeneity in detection probability.

\subsection{Conclusions}\label{sec:conclusions}
 Properly accounting for spatial autocorrelation in detection probability can mitigate bias in population size estimates. Dimension reduction of the random effects is a computationally stable technique to avoid overfitting of such complex models, but caution should be applied when choosing the aggregation scale given the trade-offs between MCMC efficiency and spatial detail. Investigators specifically interested in predicting detection probability surfaces, should choose SARE in situations where spatial autocorrelation is high and number of detections per detector is above 1. In situations where either detectability or autocorrelation is low to moderate, we recommend to use FM instead.
 









\section*{Acknowledgements}
This work was funded by the Norwegian Environment Agency (Miljødirektoratet), the Swedish Environmental Protection Agency (Naturvårdsverket), and the Research Council of Norway (NFR 286886). 

  
\section*{Supplementary material}
Additional tables and figures can be found in supplementary material
(\url{https://www.dropbox.com/s/9ur4nvyy808j2hi/HetDetSol_Supporting_Material_arxiv.pdf?dl=0}). R code for generating simulated data and data analysis are available at: \url{https://github.com/soumenstat89/HetDetSol}. 


\bibliographystyle{apalike}
\bibliography{bibliography}

\newpage

\begin{table}[H] \centering 
	\caption{\footnotesize Summary of simulated SCR data sets calculated over 100 replications for each of the 10 simulation scenarios. The last three columns show mean values.} 
	\label{table:ndets} 
	{\scriptsize\begin{tabular}{@{\hspace{0em}} l @{\extracolsep{8pt}}  l @{\extracolsep{8pt}}  r @{\extracolsep{15pt}}  r @{\extracolsep{5pt}}  r @{\extracolsep{5pt}} r @{\extracolsep{15pt}} r @{\extracolsep{5pt}} r  @{\extracolsep{5pt}} r @{\extracolsep{15pt}} r @{\extracolsep{12pt}} r
	@{\extracolsep{12pt}}  r } 
		\\[-1.8ex]\hline 
		\hline \\[-1.8ex] 
		\multirow{4}{*}{\hspace{-5pt}\begin{tabular}{l} Scenario \\ {\ } \\ {\ } \\ {\ }  \end{tabular}} & \multirow{4}{*}{\begin{tabular}{l} $\eta$\\ {\ } \\ {\ } \\ {\ }  \end{tabular}\hspace{-5pt}} &
		\multirow{4}{*}{\begin{tabular}{l} $\phi$\\ {\ } \\ {\ } \\ {\ }  \end{tabular}\hspace{-5pt}} &
		\multicolumn{3}{c}{No. of detected individuals} & \multicolumn{3}{c}{No. of detections}  &
	\multirow{4}{*}{\hspace{-12pt}\begin{tabular}{r} Detections\\ per \\ detector \\ {\ }  \end{tabular}\hspace{-5pt}} &
	\multirow{4}{*}{\hspace{-12pt}\begin{tabular}{r} Detections\\ per \\ individual \\ {\ } \end{tabular}\hspace{-5pt}} &
	\multirow{4}{*}{\hspace{-12pt}\begin{tabular}{r} Detections\\ per \\ detected \\ individual  \end{tabular}\hspace{-5pt}}\\ [0.3em]
	\cline{4-6} \cline{7-9} \\[-0.5em]
	 & & & Mean & 2.5\% & 97.5\% & Mean & 2.5\% & 97.5\% & & &  \\ 
	& & &   & Quantile & Quantile &  & Quantile & Quantile & & & \\ 
		[0.5em] \hline \\[-1ex]  
\multicolumn{10}{l}{\hspace{-5pt}Continuous}  \\ [0.5em]
$1$ & $0.1$ & $1$ & $179$ & $164$ & $197$ & $372$ & $326$ & $432$ & $0.36$ & $1.24$ & $2.08$ \\ 
$2$ & $0.1$ & $0.05$ & $163$ & $92$ & $224$ & $407$ & $127$ & $921$ & $0.40$ & $1.36$ & $2.33$ \\ 
$3$ & $0.3$ & $1$ & $229$ & $215$ & $245$ & $926$ & $843$ & $1041$ & $0.90$ & $3.09$ & $4.04$ \\ 
$4$ & $0.3$ & $0.05$ & $221$ & $186$ & $250$ & $984$ & $446$ & $1814$ & $0.96$ & $3.28$ & $4.37$ \\ 
$5$ & $0.6$ & $1$ & $244$ & $233$ & $255$ & $1610$ &  $1495$ & $1753$ & $1.57$ & $5.37$ & $6.59$ \\ 
$6$ & $0.6$ & $0.05$ & $243$ & $223$ & $257$ & $1675$ & $1106$ & $2357$ & $1.64$ & $5.58$ & $6.87$  \\  [1em]
	
\multicolumn{10}{l}{\hspace{-5pt}Categorical}  \\ [0.5em]
$7$ & $0.1$ & $1$ & $103$ & $89$ & $119$ & $140$ & $138$ & $162$ & $0.14$ & $0.47$ & $1.37$ \\ 
$8$ & $0.1$ & $0.05$ & $93$ & $82$ & $111$ & $144$ & $144$ & $166$ & $0.14$ & $0.48$ & $1.55$ \\ 
$9$ & $0.3$ & $1$ & $186$ & $170$ & $205$ & $415$ & $412$ & $468$ & $0.41$ & $1.38$ & $2.23$ \\ 
$10$ & $0.3$ & $0.05$ & $150$ & $128$ & $176$ & $422$ & $419$ & $492$ & $0.41$ & $1.41$ & $2.82$ \\   [0.5em]

	\hline\\[-0.2em] 

	\end{tabular} }
\end{table} 

\begin{table}[H] \centering 
	\caption{\footnotesize Percentage of converged replicates with respect to the top-level parameters (e.g., $N$, $\sigma$, $\eta$, $\log(\phi)$, $\pi$) for the 100 replicated SCR data sets for each of the 10 simulation scenarios tested and five fitted models. Here, `$1\times1$' indicates that the model is fitted without aggregation and `$4\times4$' refers to the level to aggregation in the random effects (Section~\ref{sec:scaling}).} 
	\label{table:nconverged} 
	{\scriptsize
	\begin{tabular}{@{\hspace{0em}} l @{\extracolsep{8pt}} r @{\extracolsep{8pt}}  r @{\extracolsep{12pt}}  r @{\extracolsep{12pt}}  r @{\extracolsep{8pt}}  r @{\extracolsep{12pt}}  r @{\extracolsep{12pt}} r @{\extracolsep{12pt}}  r @{\extracolsep{12pt}} r
	@{\extracolsep{12pt}} r} 
		\\[-1.8ex]\hline 
		\hline \\[-1.8ex] 
	\multirow{3}{*}{\hspace{-5pt}\begin{tabular}{l} Scenario\\ \\ \\ \end{tabular}\hspace{-1pt}} &
	\multirow{3}{*}{\hspace{-12pt}\begin{tabular}{l} $\eta$\\ \\ \\  \end{tabular}\hspace{-5pt}} &
	\multirow{3}{*}{\hspace{-12pt}\begin{tabular}{l} $\phi$\\ \\ \\  \end{tabular}\hspace{-5pt}} &
	
	SCR & \multicolumn{2}{c}{RE} & \multicolumn{2}{c}{SARE} & \multicolumn{2}{c}{FM} & FE \\ [0.4em]
	\cline{5-6} \cline{7-8} \cline{9-10} \\[-0.5em]
	
	& & & & $1\times1$ & $4\times4$ &  $1\times1$ & $4\times4$ & $1\times1$ & $4\times4$ &  \\[0.3em]

	\hline\\[-0.2em] 
	\multicolumn{11}{l}{\hspace{-5pt}Continuous}  \\ [0.5em]
$1$ & $0.1$ & $1$ & $100$ & $34$ & $100$ & $0$ & $18$ & 14 & 84  & $100$\\ 
$2$ & $0.1$ & $0.05$ & $100$  & $42$ & $100$ & $0$ & $67$ & 6& 86 & $100$\\ 
$3$ & $0.3$ & $1$ & $100$ & $11$ & $100$ & $0$ & $39$ & 24 & 97 & $100$ \\ 
$4$ & $0.3$ & $0.05$ & $100$ & $16$ & $100$ & $0$ & $94$ & 11 & 99 & $100$ \\ 
$5$ & $0.6$ & $1$ & $100$  & $8$ & $100$ & $0$ & $29$ & 6 & 96 & $100$\\ 
$6$ & $0.6$ & $0.05$ & $100$ & $6$ & $100$ & $0$ & $87$ & 5 & 99 & $100$ \\  [1em]
	
		\multicolumn{11}{l}{\hspace{-5pt}Categorical}  \\ [0.5em]
$7$ & $0.1$ & $1$ & $100$ & $78$ & $100$ & $0$ & $5$ & 0 & 66 & $100$ \\ 
$8$ & $0.1$ & $0.05$ & $100$ & $60$ & $100$ & $0$ & $79$ & 2 & 86 & $100$ \\ 
$9$ & $0.3$ & $1$ & $100$ & $25$ & $100$ & $0$ & $39$ & 48 & 80 & $100$ \\ 
$10$ & $0.3$ & $0.05$ & $100$ & $20$ & $100$ & $0$ & $96$ & 24 & 99 & $100$  \\   [0.5em]

	\hline\\[-0.2em] 

	\end{tabular} }
\end{table} 
 \pagebreak
\begin{table}[H] \centering 
	\caption{\footnotesize Mean MCMC efficiency of the fitted models (SCR, FE, SARE, RE and FM).  Here we report `MCMC efficiency' averaged over each top-level parameters in a model (e.g., $N$, $\sigma$, $\eta$) and over each of the converged replicates. MCMC efficiency is calculated as `ESS/MCMC run time' where the ESS (i.e., effective sample size) is based on the combined samples from the converged MCMC chains after discarding the burn-in period. The MCMC run time is calculated excluding the burn-in period. Scenarios without any converged replicates are denoted by `-'. Here, `$1\times1$' indicates that the model is fitted without aggregation and `$4\times4$' refers to the level to aggregation in the random effects (Section~\ref{sec:scaling}).}
	\label{table:MCMCeff} 
	{\scriptsize
	\begin{tabular}{@{\hspace{0em}} l @{\extracolsep{8pt}} r @{\extracolsep{8pt}}  r @{\extracolsep{12pt}}  r @{\extracolsep{12pt}}  r @{\extracolsep{12pt}}  r @{\extracolsep{12pt}}  r  @{\extracolsep{12pt}}  r @{\extracolsep{12pt}}  r @{\extracolsep{12pt}}  r @{\extracolsep{12pt}}  r } 
		\\[-1.8ex]\hline 
		\hline \\[-1.8ex] 
	
	\multirow{3}{*}{\hspace{-5pt}\begin{tabular}{l} Scenario\\ \\ \\ \end{tabular}\hspace{-1pt}} &
	\multirow{3}{*}{\hspace{-12pt}\begin{tabular}{l} $\eta$\\ \\ \\  \end{tabular}\hspace{-5pt}} &
	\multirow{3}{*}{\hspace{-12pt}\begin{tabular}{l} $\phi$\\ \\ \\  \end{tabular}\hspace{-5pt}} &
	
	SCR & \multicolumn{2}{c}{RE} & \multicolumn{2}{c}{SARE} & \multicolumn{2}{c}{FM} & FE \\ [0.4em]
	\cline{5-6} \cline{7-8} \cline{9-10} \\[-0.5em]
	
	& & & & $1\times1$ & $4\times4$ &  $1\times1$ & $4\times4$ & $1\times1$ & $4\times4$ &  \\[0.3em]
	
	\hline\\[-0.2em] 

	\multicolumn{11}{l}{\hspace{-5pt}Continuous}  \\ [0.5em]
$1$ & $0.1$ & $1$  & $1.22$ & $0.56$  & $0.54$ & - & $1.03$ & $0.07$ & $0.04$  & $1.07$\\ 
$2$ & $0.1$ & $0.05$ & $1.31$ & $0.49$ & $0.52$  & - & $0.81$  & $0.09$ & $0.05$ & $1.06$ \\ 
$3$ & $0.3$ & $1$ & $1.85$  & $0.81$ & $0.87$ & - & $1.36$  & $0.09$ & $0.06$ & $1.32$\\ 
$4$ & $0.3$ & $0.05$ & $1.92$ & $0.85$ & $0.83$ & - & $1.26$  & $0.11$ & $0.07$  & $1.31$\\ 
$5$ & $0.6$ & $1$ & $2.10$  & $0.85$ & $1.00$ & - & $1.56$  & $0.11$ & $0.06$ & $1.41$ \\ 
$6$ & $0.6$ & $0.05$ & $2.14$ & $0.90$ & $1.00$ & - & $1.41$  & $0.12$ & $0.07$ & $1.42$ \\  [1em]
	
		\multicolumn{11}{l}{\hspace{-5pt}Categorical}  \\ [0.5em]
$7$ & $0.1$ & $1$ & $0.41$ & $0.18$ & $0.23$ & - & $0.77$  & - & $0.01$ & $0.88$ \\ 
$8$ & $0.1$ & $0.05$ & $0.84$  & $0.36$ & $0.31$ & - & $0.27$  & $0.06$ & $0.04$ & $0.90$\\ 
$9$ & $0.3$ & $1$ & $1.39$ & $0.61$ & $0.59$ & - & $0.70$   & $0.08$ & $0.04$ & $1.18$ \\ 
$10$ & $0.3$ & $0.05$ & $2.31$ & $0.90$ & $1.00$ & - & $0.54$  & $0.13$ & $0.09$ & $1.15$ \\  [0.5em]

	\hline\\[-0.2em] 

	\end{tabular} }
\end{table} 

\pagebreak
\vspace*{-2.2cm}
\thispagestyle{empty}

\vspace*{-1.5cm}


 \end{document}